%% file: main.v2.tex
\documentclass[prl,aps,twocolumn, floatfix, superscriptaddress,showpacs]{revtex4}
\usepackage{amsmath,bm,graphicx}
\usepackage{amssymb}
\usepackage{amsfonts}
\newcommand{\Lp}{\Lambda^{+}}
\newcommand{\Lm}{\Lambda^{-}}
\newcommand{\Expect}{{\mathbb E}}
\newcommand{\xv}{{\mathbf x}}
\newcommand{\V}[1]{{\bf #1}}
\newcommand{\pd}[2]{\frac{\partial #1}{\partial #2}}
\newcommand{\dd}[2]{\frac{d #1}{d #2}}

\begin{document}

\title{Craig's XY--distribution and the statistics of Lagrangian power in two-dimensional turbulence}

\author{Mahesh Bandi}
\email[Corresponding Author: ]{mbandi@lanl.gov}
\affiliation{Condensed Matter and Thermal Physics Division \& Center for Nonlinear Studies, LANL, Los Alamos, NM 87545, USA}
\author{Colm Connaughton}
\email{connaughtonc@gmail.com}
\affiliation{Theoretical Division \& Center for Nonlinear Studies, LANL, Los Alamos, NM 87545, USA}
\affiliation{Centre for Complexity Science \& Mathematics Institute, University of Warwick, Coventry CV4 7AL, UK}

\date{\today}

\begin{abstract}
We examine the probability distribution function (pdf) of energy injection 
rate (power) in numerical simulations of stationary two--dimensional (2D) 
turbulence in the Lagrangian frame. The simulation is designed to mimic an 
electromagnetically driven fluid layer, a well-documented system for generating
two--dimensional turbulence in the laboratory. In our simulations, the
forcing and velocity fields are close to Gaussian. On the other hand, 
the measured PDF of injected power is very sharply peaked at zero, suggestive  
of a singularity there, with tails which are exponential but asymmetric. 
Large positive fluctuations are more probable than large negative fluctuations.
It is this asymmetry of the tails, which leads to a net positive mean value for
the energy input despite the most probable value being zero. The main features 
of the power distribution are well described by Craig's XY distribution for the
PDF of the product of two correlated normal variables. We show that the power 
distribution should exhibit a logarithmic singularity at zero and decay 
exponentially for large absolute values of the power.  We calculate the 
asymptotic behavior and express the asymmetry of the tails in terms of the 
correlation coefficient of the force and velocity. We compare the measured pdfs
with the theoretical calculations and  briefly discuss how the power pdf might 
change with other forcing mechanisms.
\end{abstract}

\maketitle
\section{Introduction}
\input intro.tex

\section{Lagrangian turbulence in two dimensions}
\input turbulence.tex

\section{Products of Normal Variables: Craig's XY Distribution}

\input productOfNormals.tex

\section{Statistics of Lagrangian Power (direct current)}
\input DCpower.tex

\section{Statistics of Lagrangian Power (alternating current)}
\input ACpower.tex

\section{Conclusions}

\input conclusion.tex

\section*{Acknowledgements}
This work was carried out under the auspices of the National Nuclear Security 
Administration of the U.S. Department of Energy at Los Alamos National 
Laboratory under Contract No.  DE-AC52-06NA25396. We would like to acknowledge
many helpful discussions with our colleagues, particularly M. Chertkov, 
R. Ecke, M. Rivera, R. Teodorescu and O. Zaboronski.

\bibliography{all}

\end{document}

%% file: intro.tex
Since turbulence is an intrinsically dissipative phenomenon, it requires an
external source of energy to sustain it. For a turbulent flow in a statistically
stationary state, the rate of injection of energy into the system from
this external energy source, the input power, is equal, {\em on average}, to
the rate of dissipation of energy by small scale viscous processes. The
fact that equality holds only on average is crucial. Since the input power is
typically calculated as a product of an external force with the fluid velocity,
both being fluctuating quantities, it is itself a fluctuating quantity with a
full statistics of its own. Locally in space or in time this injected power
need not balance the corresponding rate of dissipation. In fact it need not 
even remain positive. 

Understanding the rate of energy injection in turbulence is of considerable
importance in an engineering context, since it relates to the power required to 
overcome turbulent drag to sustain the rotation of a fan or turbine at a given 
speed in a turbulent flow (see, for example, \cite{SCH1979}). In this context,
much research focusses on the mean value of the power and how it scales with the
Reynolds number. Such measurements, focussing on the mean power,  in the 
context of Taylor--Couette flow were first performed by Lathrop et al. 
\cite{LFS1992}. Subsequent research has focussed on measuring the pdf
of the power fluctuations as well as the mean value in several
turbulent systems including von-K\'{a}rm\'{a}n flows \cite{LPF1996,TC2003}, electro-convection 
\cite{TG2003}, wave turbulence \cite{FLF2008} and turbulent convection 
\cite{AF2003}. In this latter case, it was the heat transfer rather than the 
power which was measured. Much of the recent work has focused on finding
macroscopic non-equilibrium  systems on which to test various non-equilibrium
``Fluctuation Relations'' \cite{ECM1993,GC1995,KUR1998} which have attracted 
much theoretical interest in recent years. In simple terms, these relations
express a symmetry of the probability distribution of some time-integrated
quantity associated with the dissipation or entropy production in a 
non-equilibrium system. For a review, see \cite{CCJ2006} and 
the references therein. 

All of the works cited above have focused on measurements of the statistics of
global quantities. All have observed non-trivial probability distributions,
typically with exponential tails but, to date, most of the discussions of 
the power distribution have been qualitative. In this article we use the
notion of Lagrangian turbulence as a local measure of the energy injection
into a turbulent flow. We measure the probability distribution from numerical
simulations and give a quantitative explanation of the observed distribution
in terms of Craig's XY-distribution for the product of correlated Gaussian
variables.  Although one can test the Fluctuation Relation for Lagrangian power 
fluctuations \cite{BC2007a}, this is not the purpose of the present article. We 
shall focus on the statistics of the power itself rather than it's time
integral.

The outline of the paper is as follows. We first give some background on 
turbulence in two dimensions, explaining the difference between the direct and
inverse cascades. We then give some details of our numerical simulations and
introduce the notion of Lagrangian measurements and the concept of Lagrangian
power as a diagnostic of the energy injection into a turbulent flow. We then
derive some asymptotic properties of the probability distribution of the
product of two correlated normal variables (Craig's XY-distribution). We then
return to the turbulence data and explain our measurements of the Lagrangian
power in terms of Craig's XY-distribution.

%% file: turbulence.tex
\begin{figure*}
\begin{center}
  \includegraphics[width=12cm]{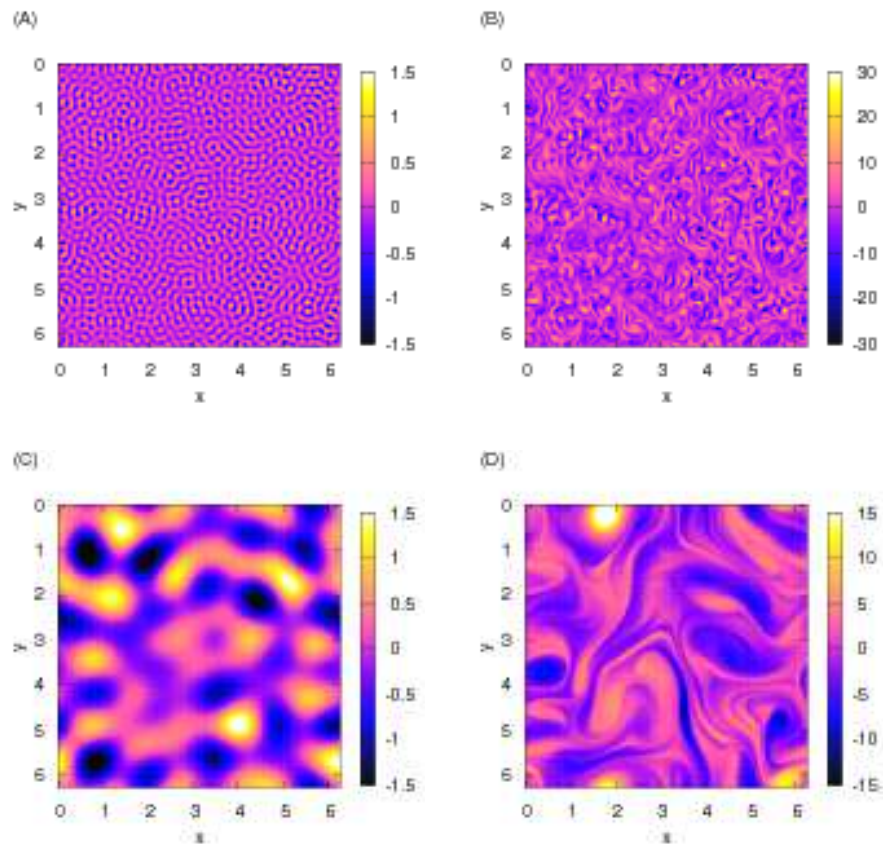}
\caption{\label{fig-snapshots} (Color online)(A) Magnetic field used to force the (256x256) inverse cascade simulation. (B) Typical vorticity snapshot 
from the inverse cascade. (C) Magnetic field used to force the (256x256) direct cascade simulation. (D) Typical vorticity snapshot 
from the direct cascade. }
\end{center}
\end{figure*}

The physics of forced turbulence in 2D differs essentially from its
three-dimensional counterpart.
This difference can be traced to the
presence of a second inviscid invariant in 2D, in addition to the energy. It is
called enstrophy. The pivotal role of the enstrophy  was first elucidated in
seminal work by Kraichnan\cite{KRA1967}, Leith\cite{LEI1968} and Batchelor
\cite{BAT1969} (KLB) which has since come to be considered as the classical
theory of turbulence in 2D. The essential insight of the KLB theory is that
simultaneous energy and enstrophy conservation requires the establishment of a
second cascade with an independent dissipation mechanism if a stationary
state is to be reached. In this dual-cascade picture, the enstrophy cascades
from the forcing scale to smaller scales whereas the energy simultaneously
cascades from the forcing scale to larger scales. This latter phenomenon,
entirely absent in 3D, is known as an inverse cascade. Viscosity dissipates
enstrophy at small scales whereas friction  between the
fluid layer and the substrate on which it moves (Ekman damping) dissipates 
energy at large scales. Assuming asymptotically
large separation of the forcing and dissipations scales for both cascades,
and applying standard Kolmogorov phenomenology\cite{FRI1995}, the KLB theory 
predicts
that the energy spectrum in the direct cascade range should scale as $k^{-3}$,
and in the inverse cascade range as $k^{-5/3}$ where $k$ is the modulus of
the wave-vector, $\vec{k}$. These spectra carry constant fluxes of
enstrophy or energy through their respective inertial ranges, that is,
ranges of scales over which forcing and dissipation are negligible.
This dual cascade theory has been shown to be in reasonable agreement with
numerical simulations, provided that a sufficiently large inertial range is
available to each cascade\cite{BOF2006}. Experimentally, both the direct and
inverse cascades have been observed\cite{BK2005}. While observing both
cascades simultaneously is rather difficult, there is broad agreement that the
KLB theory is correct asymptotically. See \cite{02KEL} for a review of
experiments.

Much research has focused on the inverse cascade in isolation since it is
responsible for the generation of large scale coherent motions in two
dimensional flow and, despite the intertwining of the two cascades in the KLB
theory, it is known to persist even when the direct cascade range is
under-developed\cite{TB2004}. Numerical experiments\cite{BCV2000} strongly
support the $k^{-5/3}$  scaling for the inverse cascade spectrum,
laboratory experiments \cite{RDE2005} are consistent and the phenomenological
KLB description of inverse energy transfer has been put in a firm quantitative
basis\cite{CEERWX2006}.

We solve the incompressible Navier--Stokes equation with Ekman term 
for the 2D velocity field, $\V{v}(\V{x},t)$:
\begin{eqnarray}
\nonumber \pd{\V{v}}{t}+\left(\V{v}\cdot \nabla\right)\V{v} +\nabla p &=& \nu\nabla^{2} \V{v} - \alpha \V{v}  + \V{f} \\
\nabla\cdot \V{v} &=&0. \label{eq-NS}
\end{eqnarray}
Here, $\V{f}(\V{x},t)$ is a forcing term (discussed below) and $p(\V{x},t)$ is
the pressure field. The parameters $\nu$ (viscosity) and $\alpha$ (Ekman 
friction coefficient) control the dissipation at small and large scales 
respectively.  As a result of using physical dissipation terms rather than
hyper-viscosity, for example, we do not develop a large inertial interval in 
our simulations, which are quite small. The modest inertial intervals in our
simulations do not pose problems. We are interested in studying energy injection 
into a turbulent flow. Unlike properties which follow from scaling arguments, this is not an asymptotic property and does not require $R\to \infty$. Nevertheless we use 
serveral different simulations at different values of $R$ to probe the robustness
of our results.  The simulations are done in a
doubly-periodic box of the size, $L=2\pi$, using a standard pseudo-spectral
solver with full dealiasing. For the inverse cascade simulations, we used
computational domains of sizes $256^2$, $512^2$ and $1024^2$ in order to investigate
the effects of varying Reynolds number. For comparison purposes, we also performed
a small simulation of the direct cascade regime at a resolution of $256^2$. 
Time integration was done using a 3rd order Runge--Kutta integrator with
integrating factors.

Unlike many numerical simulations of turbulence, the forcing term which we used
is deterministic. It was designed to model the electromagnetic forcing which
is a popular experimental method of driving turbulence in thin  fluid layers 
\cite{PT1998,RDE2005}.
The idea is to place an array of magnets underneath the fluid layer in some
particular arrangement. One then passes an electric current, which may have
some non-trivial time dependence, through the
fluid layer so that the Lorentz force produces a quasi-2D body force on the
layer. Suppose that the electric current is applied in the x-direction.
Denote it by $\V{I}(t) = I(t)\,\hat{\V{x}}$, with $\hat{\V{x}}$ being the unit vector in 
the x-direction. If $\V{B}(\V{x}) = B(\V{x})\,\hat{\V{z}}$ denotes the
(vertical component of) the magnetic field generated by the magnet array, then
the body force exerted on the fluid is
\begin{equation}
\V{f}(\V{x},t) = g\,\V{I}(t) \times \V{B}(\V{x})
\label{eq-forcing}
\end{equation}
where $g$ is a phenomenological coupling parameter which measures the strength
of the coupling of the fluid to the magnetic field. Having applied the current in
the x-direction, the force acts purely in the y-direction. In this article, 
we consider both the direct current (DC) case in which $I(t)$ is independent
of time and the alternating current (AC) case where $I(t)$ is a sinusoidal. Both
are experimentally relevant. To produce an inverse
cascade, the magnetic field is chosen to excite modes at small scales. For the
direct cascade the magnetic field is chosen to excite modes at large scales
Fig.~\ref{fig-snapshots}(A) shows the magnetic field distribution,
$B(\V{x})$,  used for forcing the inverse cascade simulations and
Fig.~\ref{fig-snapshots}(C) shows the corresponding field for the direct
cascade simulations. The field $B(\V{x})$ is generated by choosing a sum of modes
in spectral space clustered around some characteristic wavenumber. These modes
are then assigned random phases and an inverse Fourier transform taken to 
produce a spatially disordered forcing field. The characteristic forcing wavenumbers 
are around 32, 64 and 128 for the inverse cascade simulations and around 3 for the
direct cascade simulation. Note that, unlike some turbulence forcing
schemes which employ a time-varying random forcing field, the disorder in
our system is quenched. That is to say, once the initial random phases are
assigned to produce a disordered magnetic field such as those shown in 
Figs.~\ref{fig-snapshots}(A) and \ref{fig-snapshots}(C), the magnetic field
remains fixed for the duration of the simulation.

Figs.~\ref{fig-snapshots}(B) and \ref{fig-snapshots}(D) show instantaneous
snapshots of the vorticity field for the inverse and direct cascades
respectively at $256^2$ resolution. Both are in the developed turbulence regime.  The two fields
are qualitatively very different. The inverse cascade vorticity field contains
many small incoherent vortices at the scale of the forcing coexisting with
larger scale clusters of like-sign vortices.
The direct cascade vorticity field is dominated by a
smaller number of more coherent vortices, which are again comparable in
size to the forcing scale, separated by long ribbons.

\begin{figure}
\begin{center}
  \includegraphics[width=7cm]{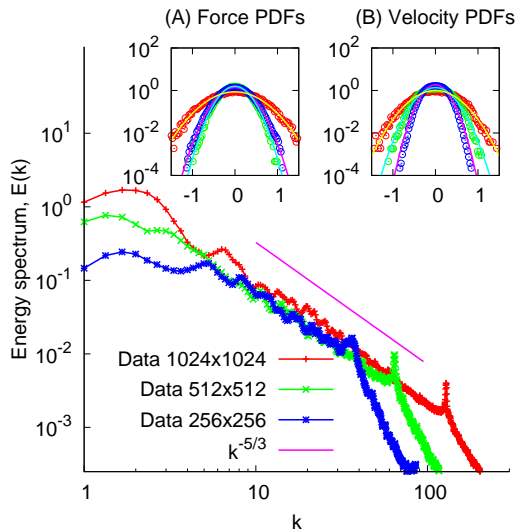}
\caption{\label{fig-spec_IC}(Color online) Snapshots of the energy spectra in the inverse cascade regime at
resolutions of $256^2$ (blue *), $512^2$ (red x) and $1024^2$ (green o). Inset (A) shows the
corresponding pdfs of the Lagrangian force and associated best fit Gaussian distributions. Inset (B) shows the
corresponding pdfs of the Lagrangian velocity and associated best fit Gaussian distributions.}
\end{center}
\end{figure}

\begin{figure}
\begin{center}
  \includegraphics[width=7cm]{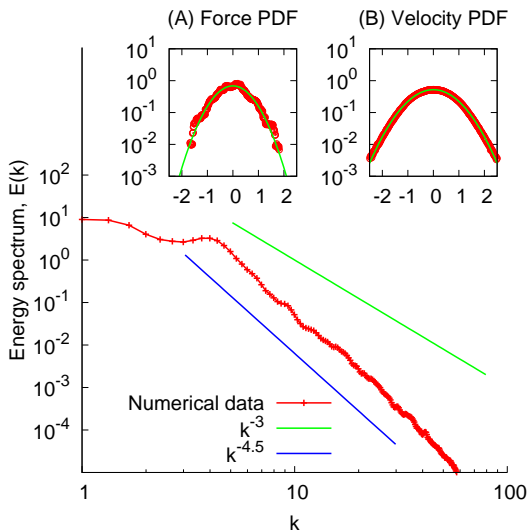} 
\caption{\label{fig-spec_DC}(Color online) Snapshot of the energy spectrum in the inverse cascade regime at
a resolution of $256^2$. Inset (A) shows the pdf of the Lagrangian force and associated best fit 
Gaussian distribution. Inset (B) shows the pdf of the Lagrangian velocity and associated best fit 
Gaussian distribution.}
\end{center}
\end{figure}

The two regimes also differ qualitatively in spectral terms. Instantaneous energy
spectra are shown in Fig.~\ref{fig-spec_IC} and Fig.\ref{fig-spec_DC}.
The inverse cascade spectrum is very close to the KLB prediction of $k^{-5/3}$,
despite the fact that the simulation does not attempt to resolve the
direct cascade range. It is accepted that the inverse cascade scaling
of $k^{-5/3}$ does not require an extensive direct cascade range despite the
fact that it is required in the theoretical argument \cite{TB2004}. On the 
other hand,
the direct cascade spectrum is much closer to $k^{-4.5}$ than $k^{-3}$. This is
again in agreement with extensive numerical and experimental investigations
of the direct cascade where it is typically found that the presence of the
coherent vortices at large scales produces spectra which are steeper than
the KLB prediction \cite{BSF2003}  and the $k^{-3}$ is only observed when very large ranges are
considered for both cascades \cite{BOF2006}.

When one talks about properties of parcels of fluid moving in turbulence, it
is natural to adopt a Lagrangian perspective. That is, one considers a set
of $N$ tracer particles, having positions, $\V{x}_i$, $i=1\ldots N$,
 which follow the fluid flow
passively. The tracer positions satisfy the advection equation:
\begin{equation}
\label{eq-lagrangianTrajectories}
\dd{\V{x}_i(t)}{t} = \V{v}(\V{x}_i(t), t)\ \ \ i=1\ldots N
\end{equation}
where $\V{v}(\V{x}, t)$ is the solution of Eq.~(\ref{eq-NS}). To study
the injection of energy into the turbulence from the forcing field, we computed 
the evolution of $100$ such Lagrangian tracers. The Lagrangian power is defined
\begin{equation}
\label{eq-powerDefn}
P(t) =  \V{v}(\V{x}_i(t), t) \cdot \V{f}(\V{x}_i(t), t).
\end{equation}
For our choice of forcing, $\V{f}$ is purely in the y-direction so that
the power is a simple product rather than a dot product, 
$P(t) =  v_y(\V{x}_i(t), t) f_y(\V{x}_i(t), t)$. We collected time series of
$v_y(\V{x}_i(t), t)$ and $f_y(\V{x}_i(t), t)$ for each tracer. Multiplying
these two together gave us a timeseries of the Lagrangian power. The velocity
in a turbulent flow, be it Lagrangian or Eulerian, is a fluctuating quantity.
Although the forcing field in Eq.(\ref{eq-forcing}) is deterministic, it is 
sampled along a random trajectory followed by the particles, so that 
$f_y(\V{x}_i(t), t)$ is also a fluctuating quantities. We computed the
empirical probability distribution functions of both $v_y(\V{x}_i(t), t)$
and $f_y(\V{x}_i(t), t)$ from our numerical data. The resulting distributions
are shown for the inverse cascade simulations in the insets of 
Fig.~\ref{fig-spec_IC} and for the direct cascade simulations in the insets
of Fig.~\ref{fig-spec_DC}. Both are close to Gaussian \cite{KF2007}. These results suggest
that the Lagrangian power can be modeled as the product of two almost Gaussian
variables which are presumably correlated to some degree 
\footnote{ Describing the pdf of injected power using two correlated normal variables has also been proposed independently by E. Falcon, S. Auma\^\i tre, C. Falc\'on, C. Laroche and S. Fauve \cite{FLF2008}}. We 
shall present measurements of the power in due course and test this assertion. First, however,
we need some understanding of the probability distribution of products of 
correlated Gaussian variables. This is addressed in the next section.

%% file: productOfNormals.tex
The product of two normally distributed random variables was first considered
by Craig \cite{CRA1937}. Consider two random variables, $x$ and $y$, which 
follow a normal bivariate distribution with means $\mu_x$ and $\mu_y$
respectively, standard deviations $\sigma_x$ and $\sigma_y$ respectively and
correlation coefficient $\rho$. Let $Z=x y$. In \cite{CRA1937}, an expression
for the moment generating function of $Z$ was derived and studied and the
distribution function, $\Pi_{xy}(Z)$, was expressed as a difference of
two integrals. For the purposes of the present work, we shall restrict ourselves
to the case of zero means, $\mu_x=\mu_y=0$, and derive $\Pi_{xy}(Z)$ directly
in a form appropriate for asymptotic analysis. 

For this case the joint distribution of $x$ and $y$ is \cite{AS1965}
\begin{equation}
\label{eq-jointDistributionxy}
\Pi(x,y) = \frac{1}{2\pi \sigma_x \sigma_y \sqrt{1-\rho^2}}\ {\rm e}^{-\frac{1}{2(1-\rho^2)}\left( \frac{x^2}{\sigma_x^2} - \frac{2\rho x y}{\sigma_x \sigma_y} + \frac{y^2}{\sigma_y^2}\right)}.
\end{equation}
To obtain the distribution of $Z$ we begin with the standard construction:
\begin{eqnarray}
\Pi_{xy}(Z) &=&  \Expect \left[ \delta(x y - Z)\right]\\
&=& \frac{1}{2\pi}\int_{-\infty}^{\infty} dw\ {\rm e}^{i Z w}\ \Expect\left[ {\rm e}^{-i x y w}\right].
\end{eqnarray}
Here, the expectation is with respect to the distribution 
(\ref{eq-jointDistributionxy}). Calculating 
$\Expect\left[ {\rm e}^{-i x y w}\right]$ is relatively simple:
\begin{eqnarray}
\nonumber \Expect\left[ {\rm e}^{-i x y w}\right] &=& \int_{-\infty}^{\infty} dx \int_{-\infty}^{\infty} dy\ \Pi(x,y)\,{\rm e}^{-i x y w}\\
\label{eq-chixy}&=& \frac{1}{2\pi \sigma_x \sigma_y \sqrt{1-\rho^2}}  \int d\xv\ {\rm e}^{-\frac{1}{2} \xv \cdot A(w) \xv},
\end{eqnarray}
where $\xv=(x,y)$ and 
\begin{equation}
A(w) = \left(\begin{array}{cc}
\frac{1}{\sigma_x^2(1-\rho^2)} & i w - \frac{\rho}{\sigma_x\sigma_y(1-\rho^2)}\\
i w - \frac{\rho}{\sigma_x\sigma_y(1-\rho^2)}& \frac{1}{\sigma_y^2(1-\rho^2)}
\end{array}\right).
\end{equation}
Evaluating the integral in Eq.(\ref{eq-chixy}) is a straightforward application 
of Gaussian integration (see, for example \cite{ID1991_vol1}):
\begin{eqnarray}
\int d\xv\ {\rm e}^{-\frac{1}{2} \xv \cdot A(w) \xv} &=& \sqrt{\frac{(2\pi)^2}{\det A(w)}}\\
&=& \frac{2 \pi}{\sqrt{(w+i\Lm)(w+i\Lp)}}
\end{eqnarray}
where
\begin{eqnarray}
\Lp &=& \frac{1}{\sigma_x\sigma_y(1+\rho)}\\
\Lm &=& \frac{1}{\sigma_x\sigma_y(1-\rho)}.
\end{eqnarray}
We are left with the following expression for the product distribution:
\begin{equation}
\label{eq-Pxy}
\Pi_{\rm xy}(Z) = \frac{\sqrt{\Lp\Lm}}{2 \pi} \int_{-\infty}^{\infty} \frac{e^{i Z w} d w}{\sqrt{(w - i \Lp)(w + i \Lm)}}.
\end{equation}
This integral cannot generally be expressed in terms of elementary functions, 
except in the case of $\rho=0$. In this case, 
$\Lm = \Lp = 1/(\sigma_x\sigma_y)$ and Eq.(\ref{eq-Pxy}) becomes proportional to
one of the integral representations of the zeroth order modified Bessel 
function of the second kind \cite{BOW1958} :
\begin{equation}
\Pi_{\rm xy}(Z) = \frac{2}{\pi \sigma_x \sigma_y} K_0(\frac{Z}{\sigma_x\sigma_y}).
\end{equation}

\begin{figure}
\begin{center}
  \includegraphics[width=7cm]{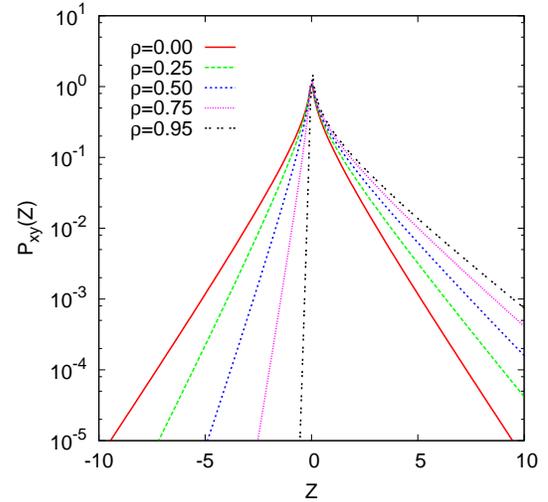}
\caption{\label{fig-Pxy}(Color online) Craig's XY-distribution for different values of $\rho$.}
\end{center}
\end{figure}

For general values of $\rho$, the integral can be evaluated numerically. Some
curves are shown for a range of positive values of $\rho$ in Fig.\ref{fig-Pxy}. 
We remark that the pdf is always peaked at zero with asymmetric exponential 
tails. The degree of asymmetry increases with increasing $\rho$. This asymmetry
reflects the fact that as the degree of correlation between $x$ and $y$ 
increases, they are increasingly likely to have the same sign. This ensures that
the probability of a positive value for the product increases with increasing
$\rho$ while the probability of a negative value decreases. For negative values
of $\rho$, or anti-correlation between $x$ and $y$, the asymmetry is in the
opposite sense.

\begin{figure}
\begin{center}
  \includegraphics[width=7cm]{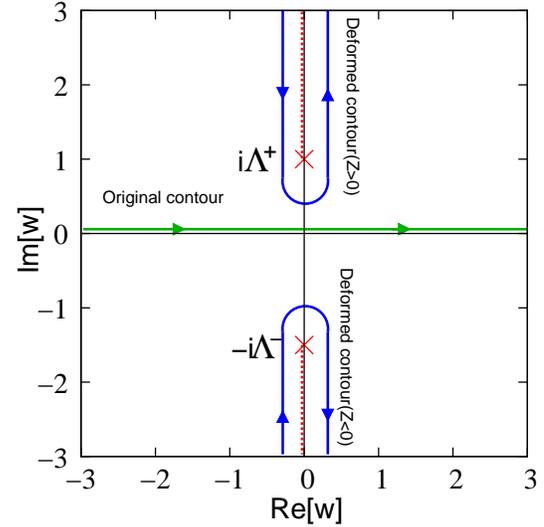}
\caption{\label{fig-analyticLandscape}(Color online) Integration contours for Eq.(1).}
\end{center}
\end{figure}

Although we cannot evaluate $\Pi_{xy}(z)$ in general, it is possible to 
calculate the asymptotic behaviour for large absolute values of $Z$ for any
$\rho$.  Let us look at the structure of the integrand in Eq.(\ref{eq-Pxy})
in the complex $w$-plane. The analytic landscape is shown in Fig.~\ref{fig-analyticLandscape}. There are two singularities on the imaginary axis at $i\Lp$ and
$-i\Lm$. The square root necessitates the introduction of a branch cut joining 
these two singularities which we take to be along the imaginary axis at 
$(i\Lp, i\infty]$ and $[-i\infty, -i\Lm)$. Depending on the sign of $Z$, we
deform the original contour of integration along the real axis into either
the upper or lower complex plane as shown in  Fig.~\ref{fig-analyticLandscape}.
Careful study of the integrand reveals that it acquires a phase difference
of $\pi$ upon going from one side of the branch cut to the other. Considering
the case $Z>0$, we may thus write:
\begin{eqnarray*}
\Pi_{\rm xy}(Z) &=& \frac{\sqrt{\Lp\Lm}}{2 \pi} \left[ \int_{i\Lp}^{i\infty} \frac{e^{i Z w} d w}{\sqrt{(w + i \Lp)(w - i \Lm)}}\right.\\
& &\left. + {\rm}e^{i\pi}  \int_{i\infty}^{i\Lp} \frac{e^{i Z w} d w}{\sqrt{(w + i \Lp)(w - i \Lm)}}\right].
\end{eqnarray*}
Rescaling $w\to i w$ and putting the two integrals together we get
\begin{equation}
\label{eq-Pxy2}
\Pi_{\rm xy}(Z) = \frac{\sqrt{\Lp\Lm}}{\pi} \int_{\Lp}^{\infty} \frac{e^{-Z w} d w}{\sqrt{(w - \Lp)(w + \Lm)}}.
\end{equation}
Changing variables, $w = \Lp + \frac{u}{Z}$ and performing some
algebra, we can write this in the form
\begin{eqnarray}
\nonumber \Pi_{\rm xy}(Z) &=& \frac{1}{\pi} \sqrt{\frac{\Lp\Lm}{\Lp+\Lm}} \frac{{\rm e}^{-\Lp Z}}{\sqrt{Z}} \int_{0 }^{\infty} du\ u^{-\frac{1}{2}} {\rm e}^{-u}\\
\label{eq-Pxy3} & & \times\left[ 1+ \frac{u}{(\Lp+\Lm) Z}\right]^{-\frac{1}{2}}.
\end{eqnarray}
To finally obtain an asymptotic series, we expand the last factor using the 
Binomial Theorem and integrate term by term (recalling the definition of
the Gamma function: $\Gamma(x) = \int t^{x-1}{\rm e}^{-t} dt$). We obtain:
\begin{eqnarray}
\label{eq-asymptoticPxy} \Pi_{\rm xy}(Z) &=& \frac{1}{\pi} \sqrt{\frac{\Lp\Lm}{\Lp+\Lm}} \frac{{\rm e}^{-\Lp Z}}{\sqrt{Z}} \\
\nonumber & & \times  \sum_{k=0}^\infty \binom{-\frac{1}{2}}{k}\Gamma(k+\frac{1}{2}) (\Lp+\Lm)^{-k} Z^{-k}.
\end{eqnarray}
Following the same approach for $Z<0$, taking the deformed contour in the lower
half plane, we obtain a very similar formula with $\Lp$ and $\Lm$ interchanged
and $Z$ replaced by $\left| Z \right|$. Noting that 
$\Gamma(\frac{1}{2})=\sqrt{\pi}$, the leading order asymptotic behaviour for
large absolute values of $Z$ is found to be:
\begin{equation}
\label{eq-PxyTails}
\Pi_{\rm xy}(Z) \sim \left\{
\begin{array}{ll}
\sqrt{\frac{\Lp\Lm}{\pi(\Lp+\Lm)}} \frac{{\rm e}^{-\Lp Z}}{\sqrt{Z}}& \mbox{$Z>0$}\\
\sqrt{\frac{\Lp\Lm}{\pi(\Lp+\Lm)}} \frac{{\rm e}^{-\Lm \left|Z\right|}}{\sqrt{\left|Z\right|}}& \mbox{$Z<0$}
\end{array} \right.
\end{equation}

The strong cusp at zero is the second striking feature of the pdfs shown in 
Fig.~\ref{fig-Pxy}. Let us briefly investigate the behaviour of the integral
in Eq.(\ref{eq-Pxy}) near $Z=0$. A fast way to compute the leading behaviour
as $Z\to 0$ is to make the change of variables $w=\Lp + u$ in 
Eq.(\ref{eq-Pxy2}), then differentiate the resulting expression with respect
to Z and apply the integrating factor $e^{\Lp Z}$, to obtain the following differential equation:
\begin{equation}
\label{eq-dPxydZ}
\dd{ }{Z}\left( e^{\Lp Z} \Pi_{xy}(Z)\right) = -\frac{\sqrt{\Lp\Lm}}{\pi} \int_0^\infty \frac{u {\rm e}^{-Z u}\ du}{\sqrt{u(u+\Lp+\Lm)}}
\end{equation}
Putting the integral on the right into Mathematica, we find that it can be
expressed in terms of a confluent hypergeometric function of the second
kind which has a tabulated Taylor expansion for small values of its
argument \cite{AS1965} :
\begin{eqnarray*}
 \int_0^\infty \frac{u {\rm e}^{-Z u}\ du}{\sqrt{u(u+\Lp+\Lm)}} &=& \frac{\sqrt{\pi}}{2 Z} \ U\left(\frac{1}{2},0,(\Lp+\Lm)\,Z\right)\\
&=&  \frac{\sqrt{\pi}}{2 (\Lp+\Lm) Z} \frac{2}{\sqrt{\pi}} + O(1).
\end{eqnarray*}
Sustituting this back into Eq.(\ref{eq-dPxydZ}) and integrating term by
term gives, to leading order in $Z$:
\begin{equation}
\label{eq-singularity}
\Pi_{xy}(Z) \sim  -\frac{\sqrt{\Lp\Lm}}{\pi}\ \log Z + O(1).
\end{equation}
Note that this simplistic approach does not provide an obvious way of 
determining the constant which arises upon integrating Eq.(\ref{eq-dPxydZ}).
Thus we do not obtain the subleading terms in the expansion near zero. For
this, a more sophisticated analysis will be required. Nevertheless, this
quick calculation is sufficient to demonstrate that the product distribution
is logarithmically singular near zero. Further, we note that, to leading order,
the singularity is symmetric about zero.

%% file: DCpower.tex
\begin{figure}
\begin{center}
  \includegraphics[width=7cm]{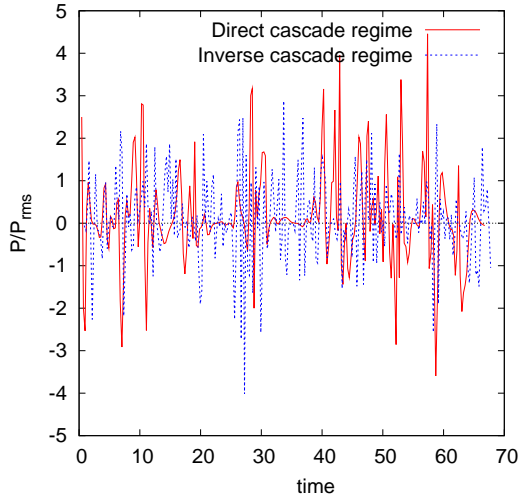}
\caption{\label{fig-powerTraces}(Color online) Typical timetraces of the Lagrangian power 
in the direct(red solid line) and inverse (blue dashed line) cascade regimes normalised by their rms values.}
\end{center}
\end{figure}

Let us now return to the Lagrangian power. We have seen from the insets of 
Fig.~\ref{fig-spec_IC} and Fig.~\ref{fig-spec_DC} that in both the inverse and
direct cascade regimes, both the velocity and force are close to Gaussian
in the Lagrangian frame. Some sample time series of the product of the two, 
the Lagrangian power, are shown in Fig.~\ref{fig-powerTraces}. We see that
both traces exhibit wild fluctuations, both positive and negative. The 
timeseries look far from Gaussian, as one might expect from the discussion
of the previous section. 

\begin{figure}
\begin{center}
  \includegraphics[width=7cm]{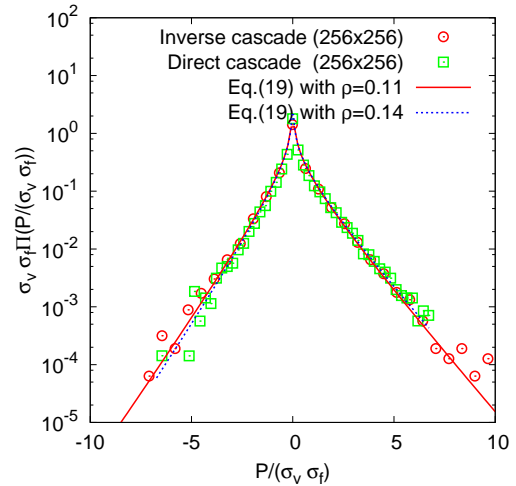}
\caption{\label{fig-powerPDF_256}(Color online) Rescaled empirical pdfs of Lagrangian power for both the inverse 
(red +) and direct (green x) cascades at $256^2$ resolution. Solid lines indicate the tails predicted 
by Eq.(\ref{eq-PxyTails}) for the measured values of the $v$-$f$ correlation coefficient, $\rho$.}
\end{center}
\end{figure}

It is clear from Eq.(\ref{eq-Pxy}) that $\sigma_x\sigma_y \Pi_{\rm xy}(\sigma_x\sigma_y Z)$
is a function of $\rho$ only. Therefore, in plotting the numerical data, we use this
rescaling together with the values of the standard deviations presented in Table~\ref{tab-data},
to collapse all of our data to similar curves.
The rescaled empirical distribution functions of the power for both
direct and inverse cascades at $256^2$ resolution are 
shown in Fig.~\ref{fig-powerPDF_256}. They almost collapse to the same
curve and are qualitatively
similar to Craig's XY-distribution shown in Fig.~\ref{fig-Pxy}. They are
strongly peaked at zero with asymmetric exponential tails. The positive tail
decays more slowly that the negative tails which gives the distribution and
net positive mean value, despite the fact that the most probable value is 
zero. The peak at zero is consistent with the logarithmic singularity 
suggested by Eq.(\ref{eq-singularity}) but such a divergence is too weak to
observe unambiguously with resolution we were able to achieve for the pdf
near zero.

\begin{figure}
\begin{center}
  \includegraphics[width=7cm]{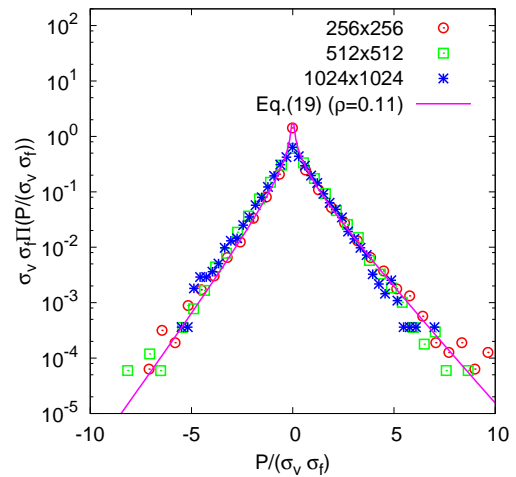}
\caption{\label{fig-powerPDF_IC_all}(Color online) Rescaled empirical pdfs of Lagrangian power for the inverse 
cascade at resolutions of $256^2$ (red o), $512^2$ (green x) and $1024^2$ (blue *) with corresponding
Reynolds numbers as noted in Table~\ref{tab-data}. The solid line indicates the tails predicted 
by Eq.(\ref{eq-PxyTails}) for the most asymmetric case ($256^2$) having $\rho=0.11$. }
\end{center}
\end{figure}

\begin{table}[ht]
\begin{tabular}{|l|l|l|l|l|l|l|}
\hline
Simulation&$v_{rms}$&$\nu$&$R$&$\sigma_f^2$&$\sigma_v^2$&$\rho$\\
\hline
IC $256^2$&0.32&$2.9\times 10^{-4}$&$6.8\times 10^3$&0.082&0.047&0.11\\
IC $512^2$&0.42&$7.3\times 10^{-5}$&$3.6\times 10^4$&0.063&0.086&0.056\\
IC $1024^2$&0.57&$1.8\times 10^{-5}$&$2.0\times 10^5$&0.21&0.16&0.027\\
DC $256^2$&1.0&$5.8\times 10^{-4}$&$1.1\times 10^4$&0.34&0.59&0.14\\
\hline
\end{tabular}
\caption{\label{tab-data}Parameters of the inverse cascade (IC) and direct cascade (DC) simulations to two significant digits. The columns are, from left to right, r.m.s. velocity, $v_{rms}$, kinematic viscosity, $\nu$, Reynolds number, $R=2\pi v_{rms}/\nu$, variance of the Lagrangian force, $\sigma_f^2$, variance of the Lagrangian velocity, $\sigma_v^2$, force-velocity correlation coefficient, $\rho$.  }
\end{table}

To make this more quantitative, we should compare the empirical distributions
with Eq.(\ref{eq-Pxy}) and Eq.(\ref{eq-asymptoticPxy}), taking $x$ to 
be the force and $y$ to be the velocity. Table~\ref{tab-data} shows the 
measured values of the correlation coefficient, $\rho$, between the
force and velocity and their standard deviations, $\sigma_f$ and $\sigma_v$.
Fig.~\ref{fig-powerPDF_256} also includes plots of the tails predicted from
Eq.(\ref{eq-PxyTails}), for the measured $\rho$ values shown in Table~\ref{tab-data}. 
In fact, the curves obtained for $\rho=0.11$ and $\rho=0.14$ are almost impossible
to distinguish at the level of convergence which we have been able to obtain
from our data. Notwithstanding the question of why the values of $\rho$ should
be so close for the two regimes, both show excellent agreement for the tails of the 
power distribution, confirming our qualitative arguments that the pdf should be close to 
Craig's XY-distribution.

Next we investigate whether Craig's XY-distribution remains a good model of the power pdf
as the Reynolds number is increased. Fig.~\ref{fig-powerPDF_IC_all} shows the rescaled pdfs
of the power obtained for the three inverse cascade simulations at resolutions of $256^2$,
$512^2$ and $1024^2$. The corresponding Reynolds numbers are shown in Table~\ref{tab-data}.
It is clear from Fig.~\ref{fig-powerPDF_IC_all} that the three pdfs almost rescale to the
same curve. There is a discernible trend towards decreasing asymmetry as the Reynolds number
is increased which is reflected in the decreasing value of the correlation coefficient, $\rho$.
This is to be expected intuitively. As the velocity becomes more turbulent, it becomes less correlated
with the forcing field. Given that we have established that the average power injected  is non-zero
due to the asymmetry of the tails, one might wonder whether the average power also decreases as $R$ increases
and the tails grow more symmetric. In fact the mean of the XY-distribution is $\sigma_v \sigma_f \rho$ so
a decreasing $\rho$ can be compensated for by increasing the variance of the velocity as $R$
increases to maintain a constant rate of energy injection. Nevertheless, it would be interesting to 
understand exactly how $\rho$ decreases with increasing $R$. The simulations presented here, while 
sufficient to identify the trend, are clearly insufficient to answer this question quantitatively.
 
One final point should be made about Fig.~\ref{fig-powerPDF_IC_all} which clearly illustrates the limitations
of this kind of modeling. While the data from the $256^2$ simulations shown in Fig.~\ref{fig-powerPDF_256}
are almost perfectly fitted by Eq.(\ref{eq-asymptoticPxy}), the $512^2$ and $1024^2$ simulations show
a smoothing out of the logarithmic cusp at zero, although the tails remain quite well captured. The 
XY-distribution only describes the power pdf as well as the underlying force and velocity distributions
can  be approximated by Gaussian distributions. Close inspection of the velocity pdfs for the
higher resolution inverse cascade simulations indicate that the pdfs of velocity become slightly flatter
than Gaussian near zero. This variation presumably accounts for the smoothing of the cusp.  One might speculate
about the physical meaning of this deviation but that is not the purpose of the present work. Indeed, in the
following section, we shall see that large deviations from Gaussianity for the forcing field can result
from simple modifications of the experimental set up.

%% file: ACpower.tex
\begin{figure}
\begin{center}
  \includegraphics[width=7cm]{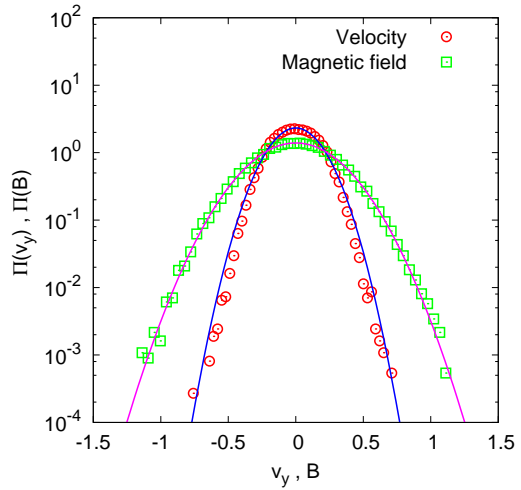}
\caption{\label{fig-acBV_IC}(Color online) Pdfs of the magnetic field (green +) and velocity (red o) in the Lagrangian frame for the inverse cascade regime with AC driving.}
\end{center}
\end{figure}

Although a Gaussian forcing field of the type observed in the direct current
simulations is often obeserved experimentally, and is often imposed
theoretically in order to simplify calculations, it is by no means true that
the forcing must be Gaussian. Thus, while the results derived above may be
applicable to a range of situations, we do not claim that they are in any
way universal. One of the simplest ways to produce a non-Gaussian
forcing is to apply AC current. We shall study this case in this section 
in somewhat less detail than the DC case. The principle aim is to compare with
the Gaussian case where we believe that the simple product distribution
discussed above provides a good model of turbulent power fluctuations. For
the AC simulations discussed below we take the current to be 
$I(t) = \cos(2\pi\omega t)$ with $\omega=10$.

For the case of AC driving, the Lagrangian force is itself a product of a 
non-trivial current and a magnetic field:
\begin{equation}
f_y(t) = I(t) B(t).
\end{equation}
Let us assume that $B(t)$ remains Gaussian.
This assumption is well supported by the numerical measurements. 
Fig.~\ref{fig-acBV_IC} shows the magnetic field and velocity distributions for
the inverse cascade regime driven by an alternating current. Both remain
close to Gaussian as in the direct current case.  So, given that
$B(t)$ is Gaussian, how is the product $f(t) = I(t) B(t)$ distributed? Of 
course, to definitively answer this question we would need to understand if and
how the magnetic field is correlated with the current. However the simplest
approximation is to assume they are uncorrelated, another
assumption which is supported by our measurements. In this case, we may treat 
$I(t)$ as a random variable generated by uniformly sampling the phase, 
$\omega t$ over the interval $[0,\frac{2\pi}{\omega}]$. If we proceed with the
calculation, this leads after some simple manipulations to the 
following expression for the pdf of $I$:
\begin{equation}
\label{eq-PI}
\Pi_I(I) = \int_{-\infty}^{\infty}\frac{d w}{2 \pi} e^{i w I} J_0(w),
\end{equation}
where $J_0(w)$ is the zeroth order Bessel function of the first kind.
This integral has a name - Weber's Discontinuous Integral (see \cite{BOW1958}, 
chap. IV), and can be evaluated as follows:
\begin{equation}
\Pi_I(I) =\left\{
\begin{array}{ll}
\frac{1}{\sqrt{1-I^2}}& \mbox{if $\left|I\right|<1$}\\
0& \mbox{if $\left|I\right|\geq 1$}
\end{array}
\right.
\end{equation}
This distribution makes sense. The current is a cosine and cannot take values
outside of the range $[-1,1]$. Furthermore, the most probable values are 
$\pm 1$ since this is where the cosine function is flatest. Let us now combine
this distribution with an assumed Gaussian distribution for $B(t)$ with mean
zero and variance $\sigma_B$. Doing some lengthy, but elementary, calculations 
in the spirit of our derivation of $\Pi_{xy}(Z)$, we obtain the following 
proposed distribution for the Lagrangian force in the case of AC driving:
\begin{equation}
\label{eq-Pfac}
\Pi_f^{\rm (ac)}(F) = \frac{1}{2\pi}\sqrt{\frac{2}{\pi \sigma_B^2}} e^{-\frac{F^2}{4 \sigma_B^2}}\ K_0\left(\frac{F^2}{4 \sigma_B^2}\right).
\end{equation}

\begin{figure}
\begin{center}
  \includegraphics[width=7cm]{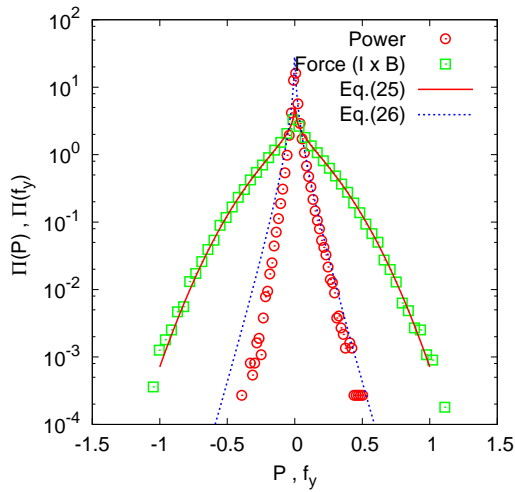}
\caption{\label{fig-acFP_IC}(Color online) Pdfs of the Lagrangian force (green o) and power
(red +) for the inverse cascade regime with AC driving.
}
\end{center}
\end{figure}

This formula is compared with the measured Lagrangian force distribution
in Fig~\ref{fig-acFP_IC}. We see that the alternatic current strongly modifies
the distribution of the force from the Gaussian distribution measured in the
direct current case. However, this modification is correctly captured by
Eq.(\ref{eq-Pfac}).

Finally, one may ask if the distribution of the power may be calculated in this
case. Here one encounters a problem. We have learned that the correlation
coefficient between the force and the velocity controls the degree of asymmetry
of the power pdf. In the direct current case, where both the force and
 the velocity are Gaussian, the bivariate normal distribution, 
Eq.(\ref{eq-jointDistributionxy}), provides a very reasonable way to correlate
the two. In the alternating current case, where the force follows the 
distribution of Eq.(\ref{eq-Pfac}), it is less obvious how to correlate
the two. In the absence of a more thorough analysis addressing this issue,
we give here the result for the uncorrelated case only. If the force
follows the distribution Eq.(\ref{eq-Pfac}) with standard deviation
$\sigma_B$  for the magnetic component and the velocity is normally 
distributed with standard deviation, $\sigma_v$, then some lengthy
calculations show that the uncorrelated product follows the 
distribution
\begin{equation}
\Pi^{\rm (ac)}(P) = \frac{1}{\pi \sigma_B \sigma_v} \int_{-\infty}^{\infty} \frac{d w}{w} e^{-\left(\frac{P^2}{2 \sigma_v^2 w^2}+\frac{w^2}{4 \sigma_B^2} \right)}
K_0\left(\frac{w^2}{4 \sigma_B^2}\right)
\end{equation}
This distribution is plotted with the measured values of $\sigma_B$ and 
$\sigma_v$ in  Fig~\ref{fig-acFP_IC} against the empirical distribution of the
Lagrangian power. As expected, it completely fails to capture the asymmetry
of the power distribution. Nevertheless it is sufficiently close to 
convince us that the heuristic arguments presented here are correct and that
some more careful calculations might be worth doing to attempt to incorporate
the correlation between $v$ and $f$ in a reasonable way. Furthermore,
the measured distribution for the power looks qualitatively very
similar to the distributions measured for the direct current case. This is
despite the fact that the force distributions are so different in the two cases.
This suggests that some qualitative features of the distribution are more 
universal than one might expect from the very particular calculations performed 
above for products of Gaussian variables. These questions may be worth further 
investigation in the future.

%% file: conclusion.tex
We have characterised the injection of energy into a two-dimensional flow by 
measuring the injected power in the Lagrangian frame. The measured distribution
is very sharply peaked at zero with asymmetric exponential tails. It can 
be understood as arising from a simple product of correlated almost-Gaussian
variables (the force and the velocity).  From our analysis, we suggest
that the power distribution is actually logarithmically singular at zero 
although this is too weak of a divergence to distinguish convincingly in 
our numerical data. We derived an asymptotic expression
for the degree of asymmetry of the tails of the power distribution which 
depends only on the correlation coefficient of the force and velocity. We
concluded by comparing the results for a case, alternating current instead
of direct current forcing, for which the observed force distribution is
far from Gaussian. The qualitative features of the power pdf remain unchanged
although it will be hard to quantify the measurements as easily as in
the Gaussian case.